\documentclass[twocolumn]{IEEEtran}

\usepackage{cite}
\usepackage{gensymb}
\usepackage{amsmath,amssymb,amsfonts}
\usepackage{bm}
\usepackage{amsthm}
\usepackage{textcomp}
\usepackage{graphicx}
\usepackage{algorithm}
\usepackage{algpseudocode}

\usepackage{multirow}
\usepackage{booktabs}
\usepackage{siunitx}
\usepackage{graphicx}
\usepackage{textcomp}
\usepackage{xcolor}
\usepackage{comment}
\usepackage{svg}
\usepackage{subcaption}

\usepackage[margin=16.3mm,top=19.1mm,bottom=26.4mm]{geometry}

\usepackage[normalem]{ulem} 
\usepackage{cancel}
\usepackage{mathtools, cuted}

\usepackage{tikz}
\usetikzlibrary{shapes,arrows,positioning}
\usetikzlibrary{3d}

  {\proof}{\proofend}

\newcommand{\qa}{{\bf a}}

\newcommand{\qc}{{\bf c}}

\newcommand{\qg}{{\bf g}}
\newcommand{\qh}{{\bf h}}

\newcommand{\qp}{{\bf p}}

\newcommand{\qw}{{\bf w}}
\newcommand{\qx}{{\bf x}}

\DeclareMathOperator*{\argmin}{arg\,min}

\newcommand{\ps}{{\qp_{s}}} 
\newcommand{\pmr}{{\qp_{m}}} 
\newcommand{\pkr}{{\qp_{k}}} 

\newcommand{\gsk}{{\qg_{s,k}}} 

\newcommand{\wsk}{{\qw_{s,k}}} 

\newcommand{\as}{{a_{s}}} 
\newcommand{\ats}{{\tilde{a}_{s}}} 

\newcommand{\asp}{{a_{s'}}} 

\newcommand{\nx}{{N_{s,x}}} 
\newcommand{\ny}{{N_{s,y}}} 
\newcommand{\ns}{{N_s}} 

\newcommand{\sus}{{\sum\nolimits^{S}_{s=1}}}
\newcommand{\summ}{{\sum\nolimits^{M}_{m=1}}}
\newcommand{\suml}{{\sum\nolimits^{L}_{l=1}}}

\newcommand{\susp}{{\sum\nolimits_{s'=1}^{S}}}

\newcommand{\ID}{{\mathsf{ID}} }

\newcommand{\PIDC}{{\Psi^{\mathsf{\,c}}_{\ID,l}(\boldsymbol{\Omega}^{\mathsf{ID}}\!\!,\qa)}} 

\newcommand{\PIDNC}{{\Psi^{\mathsf{\, nc}}_{\ID,l}(\boldsymbol{\Omega}^{\mathsf{ID}}\!\!,\boldsymbol{\Omega}^{\mathsf{EH}}\!\!,\qa)}} 

\newcommand{\OID}{{\boldsymbol{\Omega}^{\mathsf{ID}}}} 
\newcommand{\OEH}{{\boldsymbol{\Omega}^{\mathsf{EH}}}} 
\newcommand{\tOmgID}{\tilde{\boldsymbol{\Omega}}^{\mathsf{ID}}} 
\newcommand{\tOmgEH}{\tilde{\boldsymbol{\Omega}}^{\mathsf{EH}}} 

\newcommand{\tOmgEHt}{\tilde{\boldsymbol{\Omega}}^{\mathsf{EH},(t)}} 
\newcommand{\tOmgIDt}{\tilde{\boldsymbol{\Omega}}^{\mathsf{ID \!,(t)}}} 

\newcommand{\OmgIDst}{{\boldsymbol{\Omega}}^{{\star}\mathsf{ID}}} 
\newcommand{\OmgEHst}{{\boldsymbol{\Omega}}^{{\star}\mathsf{EH}}} 

\newcommand{\osm}{{\Omega_{s,m}^{\mathsf{EH}}}}
\newcommand{\osl}{{\Omega_{s,l}^{\mathsf{ID}}}}

\newcommand{\ME}{{\mathbb{E}}} 

\title{Near-Field SWIPT Using XL-MIMO:\\ Power Allocation and Subarray Activation}
\author{\IEEEauthorblockN{Muhammad Zeeshan Mumtaz, Mohammadali Mohammadi, Hien Quoc Ngo, and Michail Matthaiou}\\
\IEEEauthorblockA{{Centre for Wireless Innovation (CWI), Queen’s University Belfast, U.K.} \\
Email:\{mmumtaz01, m.mohammadi, hien.ngo, m.matthaiou\}@qub.ac.uk}
\thanks{

This work was supported by the U.K. Engineering and Physical Sciences Research Council (EPSRC) (grants No. EP/X04047X/1 and EP/X040569/1). The work of  H. Q. Ngo
 was supported by the U.K. Research and Innovation Future
Leaders Fellowships under Grant MR/X010635/1, and a research grant from the Department for the Economy Northern Ireland under the US-Ireland R\&D Partnership Programme. This work was supported by the European
Research Council (ERC) under the European Union’s Horizon 2020 research
and innovation programme (grant agreement No. 101001331). }
\thanks{M. Z. Mumtaz is also with the College of Aeronautical Engineering, National University of Sciences \& Technology (NUST), Pakistan, (email: zmumtaz@cae.nust.edu.pk).}
}
\date{}

\begin{document}

\maketitle

\begin{abstract}
    This paper investigates the simultaneous wireless information and power transfer (SWIPT) capability of a modular extremely large multiple-input multiple-output (XL-MIMO) system, in the context of power consumption (PC) efficiency. The network users are divided into two functional categories: information decoding (ID) users and energy harvesting (EH) users. Non-stationary near-field channels are considered whilst the users are located in spatially distinct visibility regions (VRs). We formulate a two-tier joint optimization problem to minimize the PC, taking into account the power allocation (PA) for ID and EH users, along with the activation of constituent XL-MIMO subarrays. This complicated mixed-integer problem is transformed into more tractable formulations and  efficient algorithms are proposed for solving them.  The numerical results demonstrate that the overall PC of the XL-MIMO system for the proposed method is reduced by more than $60\%$ in comparison to the benchmark scheme of equal PA with full subarray activation (SA) and $30\%$ against the case of optimized PA with full SA, while satisfying the quality-of-service  (QoS) constraints on both the downlink rate of the ID users and harvested energy at the EH users.
\end{abstract}
\vspace{-0.5em}

\vspace{-0.5em}
\section{Introduction}
In near-field communications, XL-MIMO systems enhance the SWIPT efficiency by leveraging spherical wavefronts for precise energy and information beamfocusing~\cite{Haiquan}. On the other hand, the concept of VR limits each user’s interaction to a specific subset of antennas, thereby enabling strategic resource allocation and energy-efficient SWIPT through energy focusing and interference reduction designs~\cite{cui2,Xinrui,kangda}.

Contemporary research in near-field SWIPT underscores an increasing emphasis on enhancing the energy efficiency in the forthcoming sixth generation (6G) wireless networks. 
In this context, a comprehensive review in \cite{Haiyang_Shlezinger} explored the deployment challenges and benefits of near-field wireless power transfer (WPT) in tandem with information transfer for future 6G Internet-of-Everything (IoE) applications, emphasizing the importance of beamfocusing in the near-field to enhance the WPT efficiency and avoid energy pollution.
In~\cite{Zhang:JSAC:2024}, the authors considered joint beam scheduling and PA for SWIPT in a mixed near- and far-field channel scenario, optimizing the EH efficiency for near-field receivers. This mixed-field approach leverages the XL-array technology to allocate dedicated power resources towards near-field EH users, while ensuring minimal interference for the ID users located in the far-field region. Zhang \textit{et al.}~\cite{Zhang:IoT:2024} proposed a SWIPT framework for near-field communication networks. By employing a hybrid beamforming strategy for transmit power minimization, the analog beamformer and baseband digital information beamformers were jointly optimized using a two-layer penalty-based algorithm. However, these studies do not address the effects of non-stationary channels, particularly given the large physical dimensions of XL-MIMO arrays, where ID and EH users are primarily served by specific portions of the array.

Given the limitations of system non-stationarities identified in the literature, operating all subarrays is not energy efficient. This motivates our research on joint optimization of PA and selective SA in SWIPT-enabled XL-MIMO systems. The main contributions of this paper are as follows:
\begin{itemize}
    \item We devise a joint optimization technique for the overall PC minimization problem in a SWIPT based XL-MIMO system, over the PA for both the ID and EH users as well as the SA, while satisfying the QoS requirements for both groups of ID/EH receivers.
    \item We split the primary optimization problem into two sub-problems: (1) PA routine, (2) SA routine. For the PA routine, the Douglas-Rachford (DR) splitting based alternating
    direction method of multipliers (ADMM) algorithm from \cite{pontus} is utilized to optimize the PA variables for both ID and EH users simultaneously. Whereas, the SA routine parametrizes the SA variables using a surrogate auxiliary function of the optimized PA coefficients.
    \item Our numerical results confirm a significant reduction in the overall PC of the envisaged XL-MIMO system using the proposed optimization algorithm with VR based SA, while meeting both the downlink (DL) rate and EH power  requirements. 
\end{itemize}

\emph{Notations:} We use bold lower case letters to denote vectors. The superscripts $\!(\cdot)^{\rm{*}}\!$ and $\!(\cdot)^{\rm{T}}\!$ denote the conjugate and transpose of a matrix, respectively; $\boldsymbol{1}_N$ denotes an all-one vector of size $N$; $\| \cdot \|$ returns the norm of a vector. Finally, $\mathbb{E}\{\cdot\}$ denotes the statistical expectation.

\vspace{-1em}
\section{System Model}
Consider a near-field SWIPT network underpinned by a modular XL-MIMO array. This XL-MIMO consists of $S$ uniform planar subarrays (UPSAs). Each $2$-dimensional subarray is equipped with $\ns$ antenna elements, consisting of $\nx$ elements along the $x$-axis and $\ny$ elements along the $y$-axis. These subarrays are serving $K$ number of single-antenna users, with $L$ ID users located in $\mathrm{{V}_{\!ID}}$ VRs and $M=K-L$ EH users in separated $\mathrm{{V}_{\!EH}}$ VRs. 
All user devices are located in the radiative near-field region of the XL-MIMO array, which means that the distances between the XL-MIMO array and users are shorter than the one-tenth of Fraunhofer array distance $d_{FA}$ \cite{Ramezani2024}. This physical parameter is defined as $d_{FA}=2 D^2 (S \ns)/\lambda$, where $D$ is the largest dimension of each antenna element and $\lambda$ is the wavelength of the carrier frequency. Within this radiative zone, the transmitted electromagnetic (EM) waves have a spherical wavefront nature with finite beam depth\cite{Ramezani2024}.

The free-space near-field radiating channel $\qg_{s,k} \in \mathbb{C}^{\ns \times 1}$ from the subarray $s$  to the user $k$ at the location $\pkr= (x_k,y_k,z_k)$, follows the cosine pattern model \cite{Haiquan}, given as
\begin{equation}\label{eq:channel}
    \gsk = \dfrac{\lambda}{4 \pi \lVert \pkr -\ps \rVert}\sqrt{F(\theta_{s,k})} \qh_{s,k} ,
\end{equation}
where
\vspace{-0.5em}
\begin{align*}
    \qh_{s,k}\!\!= \!\!\Big[e^{\!-j k \lVert \pmr\! -\! \qp_{s_{1,1}} \! \rVert},\!e^{\!-j k \lVert \pmr\! - \!\qp_{s_{1,2}}\!  \rVert}\!,\!\hdots\!,\! 
e^{\!-j k \lVert \pmr \!-\! \qp_{s_{N_x,N_y}}\!  \rVert}\Big]^T\!,
\end{align*}
represents the channel vector from that particular subarray $s$ whose antenna elements are positioned at the locations $(x_{s_{1,1}}, y_{s_{1,1}},0), (x_{s_{1,2}},y_{s_{1,2}},0), \hdots, (x_{s_{x,y}},y_{s_{{x,y}}}, 0)$; $e^{-j 2\pi\lvert \pkr -\ps \rvert/\lambda}$ denotes the phase attributed to the distance covered by the EM wave from $\ps$ to $\pkr$; $F(\theta_{s,k})$ is the cosine radiation profile of each antenna element for the boresight gain $b$, defined as~\cite{Haiquan}
\begin{align}
    F(\theta_{s,k}) &= 
    \begin{cases}
        2 (b+1) \,\text{cos}^{b}(\theta_{s,k}) \quad  \theta_{s,k} \in [0,\pi/2],\\
        0 \hspace{3.5cm} \text{otherwise}.
    \end{cases}
\end{align}

We consider a fully-digital array configuration with a dedicated radio-frequency (RF) chain for each antenna element. Furthermore, we utilize the maximum ratio transmission (MRT) technique to evaluate the precoding vector for the user $k$ from the subarray $s$ as $\wsk= \gsk/\lVert\gsk\rVert_2$. Note that MRT precoding has been established as the optimal precoding method for WPT, especially with a large number of antennas~\cite{cite:MRT_for_HE:Almradi}.
The transmitted signal by the subarray $s$ is given as
\begin{equation}
    \qx_s \! =\! \as \bigg(\suml \sqrt{\osl} \qw^*_{s,l} x_{l} \! + \! \summ\!   \sqrt{\osm} \qw^*_{s,m} e_{m} \bigg),
\end{equation}
where $x_{l} \in \mathbb{C}$ is the information bearing signal for ID user $l$, and $e_{m}$ is the normalized zero-mean pseudo-random energy signal, both satisfying $\ME\{\lvert x_{l} \rvert^2\}=\ME\{|e_{m}|^2\}=1$. Moreover, $\as$ is a binary variable, which indicates whether a particular subarray $s$ is functionally active or not. It is mathematically represented as
\begin{align}
    \as &= 
    \begin{cases}
        1 \quad \quad \text{if subarray $s$ is switched on},\\
        0 \quad \quad \text{if subarray $s$ is not switched on}.
    \end{cases}
\end{align}

Meanwhile, $\osl$ and $\osm$ are the power allocation coefficients for the ID and EH users which are selected to satisfy the power constraint at each subarray, such that
\begin{equation}
    \mathbb{E}\left\{\lVert \qx_s \rVert^2 \right\}= \suml \osl + \summ  \osm   \leq P_s,
\end{equation}
where $P_s= \ns P_{\mathrm{et}}$ is the maximum subarray transmit power, while $P_{\mathrm{et}}$ is the maximum transmit power of each antenna element. The overall XL-MIMO transmit power of all modular subarrays is $P_t = S P_s$. Consequently, the overall power consumed by the whole XL-MIMO array during the DL phase  can be mathematically represented as~\cite{Jun_Zhang,Asaad}
\begin{align}\label{eq:power_consumed}
    P_C(\OID\!\!,\OEH\!\!,\qa&)\! =\! \sus\! \as \bigg[\frac{1}{\varsigma} \bigg( \!\suml\! \, \osl\!+ \!\summ \, \osm \bigg)  \nonumber\\ 
    & \quad + 2 P_{\mathrm{syn}} + N_s P_{\mathrm{ct}}\bigg],
\end{align}
where $\OID\triangleq \{\osl\}_{l=1,\ldots,L}^{s=1,\ldots,S}$;  $\OEH\triangleq \{\osm\}_{m=1,\ldots,M}^{s=1,\ldots,S}$; $\qa=\{a_s\}_{s=1,\ldots,S}$;  $\varsigma$ is the efficiency of the power amplifier; $P_{\mathrm{syn}}$ is the power consumed by the frequency synthesizer of each subarray; $P_{\mathrm{ct}}$ is the circuit power consumed by each antenna RF chain.

The received signal at the ID user $l$ and EH user $m$, can be respectively expressed as
\begin{subequations}
    \begin{align}
        r_{l} & = \sus \qg^T_{s,l} \qx_s + n_{l}, \quad \,\,\,\,\, \forall \, l=1,2, \hdots, L, ~\label{eq:rl} \\
        r_{m} &= \sus \qg^T_{s,m} \qx_s+ n_{m}, \quad  \forall \, m=1,2, \hdots, M,
    \end{align} 
\end{subequations}
where $n_l, n_{m} \sim \mathcal{CN}(0, \sigma^2_l)$ denotes the additive white Gaussian noise (AWGN) at ID user $l$ and EH user $m$, respectively. 

By using~\eqref{eq:rl}, the DL achievable spectral efficiency of ID user $l$ can be written as
\vspace{-0.3em}
\begin{align}~\label{eq:DL_rate}
R_l(\OID\!\!,\OEH\!\!,\qa) &=  \log_2 \!\Bigg(1+ \dfrac{\PIDC}{\PIDNC}\Bigg),
\end{align}
where
\begin{subequations}~\label{eq:DL_rate2}
    \begin{align}
        \PIDC &\triangleq\sus \as \osl  \lvert\qg^T_{s,l} \qw^*_{s,l} \rvert^2,\\
        \PIDNC &\triangleq\sum\nolimits_{l' \neq l} \sus  \as \Omega_{s,l'}^{\ID}  \lvert\qg^T_{s,l} \qw^*_{s,l'} \rvert^2\nonumber \\
        &\hspace{-4em} + \summ \sus  \as \osm  \lvert\qg^T_{s,l} \qw^*_{s,m} \rvert^2 + \sigma^2_{l},
    \end{align}
\end{subequations}
where $\PIDC$ is the desired coherent signal intended for user $l$ from all subarrays and $\PIDNC$ is the non-coherent signal received at the same user which includes the undesired effects of the interference from other ID user signals and DL EH transmitted power,  and also the AWGN noise. 

The input energy at the reception antenna of EH user $m$ is mainly composed of two RF signal components: (1) the received energy from the transmitted power to EH users, $I^{\mathsf{EH}}_m (\OEH\!\!,\qa)$, (2) the received energy from the information signal transmitted towards ID users, $I^{\mathsf{ID}}_m  (\OID\!\!,\qa)$. Therefore, the input energy can be expressed as
\begin{equation}\label{eq:harvested_power}
    I_m (\OID\!\!,\OEH\!\!,\qa) = I^{\mathsf{EH}}_m (\OEH\!\!,\qa) + I^{\mathsf{ID}}_m  (\OID\!\!,\qa),
\end{equation}
where
\begin{align}
    I^{\mathsf{EH}}_m (\OEH\!\!,\qa)& = \sum\nolimits^M_{m'=1} \Big\lvert \sus \, \as \sqrt{\Omega^{\mathsf{EH}}_{s,m'}} \qg^T_{s,m} \qw^*_{s,m'} \Big\rvert^2 \nonumber\\
     &\hspace{-3em}=\!\!\sum\nolimits^M_{m'=1} \sus \susp \as \asp \sqrt{\Omega^{\mathsf{EH}}_{s,m'}\Omega^{\mathsf{EH}}_{s',m'}} \Upsilon_{s,s'\!,m, m'},\nonumber
\end{align}
and
\begin{align}
    I^{\mathsf{ID}}_m  (\OID\!\!,\qa) & = \suml \Big\lvert \sus \, \as \sqrt{\osl}  \qg^T_{s,m} \qw^*_{s,l} \Big\rvert^2 \nonumber\\
    &\hspace{-2em}= \suml  \sus \susp \as \asp \sqrt{\Omega^{\mathsf{ID}}_{s,l}\Omega^{\mathsf{ID}}_{s',l}} \Upsilon_{s,s'\!,m, l},\nonumber
\end{align}
where $\Upsilon_{s,s'\!,m, m'}=\qg^T_{s,m} \qw^*_{s,m'} \qg^T_{s',m} \qw^*_{s',m'}$ and $\Upsilon_{s,s'\!,m, l}=\qg^T_{s,m} \qw^*_{s,l}\qg^T_{s',m} \qw^*_{s',l}$.  It is important to note that noise has not been considered as a contributing factor in the EH process \cite{Zhang:JSAC:2024}.

On the account of a realistic SWIPT application of this XL-MIMO system, 
a practical nonlinear EH (NL-EH)  model is considered \cite{Boshkovska,Nezhadmohammad}. Therefore, the harvested energy at user $m$ is given by
\begin{align}
    I^{\mathsf{NL}}_m (\OID\!\!,\OEH\!\!,\qa)= \dfrac{\psi_m(I^{\mathsf{NL}}_m (\OID\!\!,\OEH\!\!,\qa)) - \zeta_{max}\varphi}{1-\varphi},
\label{eq:non_linear_harvested_power}
\end{align}
where  $\psi_m(I^{\mathsf{NL}}_m (\OID\!\!,\OEH\!\!,\qa)) \!=\! \zeta_{max}/(1+ e^{-a [I^{\mathsf{NL}}_m (\OID\!\!,\OEH\!\!,\qa)-b]})$ is the traditional logistic function of the received EH power, where $a, b$ are circuit related parameters and $\zeta_{max}$ is the maximum output DC power at each user. Moreover, $\varphi=1/{(1+e^{ab})}$ is a constant to guarantee a zero input/zero-output response.

\vspace{-1em}
\section{Joint Optimization Process}
In this section, we formulate and solve an optimization problem that optimally allocates the powers and controls the subarray activation to minimize the overall PC of the XL-MIMO array given in \eqref{eq:power_consumed} while maintaining certain DL rate for each ID user and harvested energy for each EH user. The optimization problem can be formulated mathematically as
\begin{subequations}
    \begin{align}
        \mathcal{P}_1:&\min_{\{\OID\!,\OEH\!,\qa\}}\!   \,\, P_C(\OID\!\!,\OEH\!\!,\qa)\\
        \hspace{0 em} \text{s.t} ~ \,\, & \,\,  R_l(\OID\!\!,\OEH\!\!,\qa) \!\geq\! R^{\mathsf{EA}}_{th},~\forall l=1,\ldots,L, \label{eq:optimization_1_C1} \\
        & \,\, I^\mathsf{NL}_m(\OID\!\!,\OEH\!\!,\qa) \!\geq\! I^{\mathsf{EA}}_{th},~\forall m=1,\ldots,M,\label{eq:optimization_1_C2}\\
        & \,\,\sus\! \as\left(\suml  \osl \! +\! \summ\!  \osm\! \right)  \leq P_t,\label{eq:optimization_1_C3}\\
        & \,\, \suml \osl + \summ  \osm  \leq P_{s},\label{eq:optimization_1_C4}\\
        & \,\,  0 \leq \osl,\,\,0 \leq \osm, \quad \as \in \{0,1\} \label{eq:optimization_1_C5},
    \end{align}    
\end{subequations}

\hspace{-1em}where $R^{\mathsf{EA}}_{th}=R_l(\OID\!\!,\OEH\!\!,\boldsymbol{1}_S)$ and $I^{\mathsf{EA}}_{th} \triangleq  I^\mathsf{NL}_m(\OID\!\!,\OEH\!\!,\boldsymbol{1}_S)$ are the QoS requirements for  ID user $l$ and EH user $m$, respectively. Considering a fair comparison based optimization criterion, the choice of these QoS thresholds is based on the equal PA and full array activation (\textbf{EA-FA}) case. This means that we minimize the system PC while ensuring the DL rate and EH power levels achieved in \textbf{EA-FA} case.

It should be noticed that constraints \eqref{eq:optimization_1_C1} and \eqref{eq:optimization_1_C2} limit the minimum DL SE for each user $l$, and harvested energy for each user $m$ to the case of equal PA denoted as $\Omega^{\mathsf{EA}}_{s,l}$ and $\Omega^{\mathsf{EA}}_{s,m}$ with full array activation $\as = 1$. On the other hand, the constraint \eqref{eq:optimization_1_C3} imposes an upper bound on the sum transmitted power of active subarrays, with a restriction to the combined transmission power available at the overall XL-MIMO array, while satisfying the power constraint \eqref{eq:optimization_1_C4} at individual subarrays.

It can be observed that the optimization problem $\mathcal{P}_1$ has mixed integer non-convex nature, with two second-order cone (SOC) constraints i.e., \eqref{eq:optimization_1_C1} and \eqref{eq:optimization_1_C2}. The conic reformulation of the problem $\mathcal{P}_1$ can be represented as:
\begin{subequations}\label{eq:optimization_2}
    \begin{align}
        \!\mathcal{P}_2:&\min_{\{\OID,\OEH,\qa\}}   \,\, P_C(\OID\!\!,\OEH\!\!,\qa) \label{eq:objective_function_2}\\
        \!\!\! \! \text{s.t} \quad  & \!\!  
        \lVert \PIDNC \rVert_2 \!\leq\! \Xi(R^{\mathsf{EA}}_l)\lVert \PIDC\rVert_2,\!\! \label{eq:optimization_2_C1} \\
        & \!\!\lVert I_m(\OID\!\!,\OEH\!\!,\qa)\rVert_2 \geq\Lambda (I^{\mathsf{NL\!-\!EA}}_m),\label{eq:optimization_2_C2}\\
        & \!\!\sus \as\left(\suml  \osl + \summ  \osm \right)  \leq P_t,\label{eq:optimization_2_C3}\\
        & \!\! \suml \osl + \summ  \osm  \leq P_{s},\label{eq:optimization_2_C4}\\
        & \!\!  0 \leq \osl,\,\,0 \leq \osm, \quad \as \in \{0,1\},\label{eq:optimization_2_C5}
    \end{align}    
\end{subequations}
where $\Xi\big(R^{\mathsf{EA}}_l\big)=1/\Big(2^{R^{\mathsf{EA}}_{th}}-1\Big)^{\! 1/2}$, and
$$\Lambda (I^{\mathsf{NL\!-\! EA}}_m)\!\!=\!\! \Bigg(\dfrac{1}{a}\log\Big[1-\dfrac{\zeta_{max}}{(1-\varphi)I^{\mathsf{EA}}_{th}+\zeta_{max}\varphi}\Big]+b\Bigg)^{\! 1/2}.$$ 

The set of optimization variables in the mixed-integer optimization problem $\mathcal{P}_2$ comprises of continuous PA variables ($\osl,\osm$) and binary SA variables $\{\as\}$. We can now devise a two-tiered iterative optimization framework presented in \textbf{Algorithm \ref{alg:opt_process}}, in which we decouple the binary SA variables by using a parametric transformation based on the PA variables. We will first discuss the optimization process of the PA sub-problem, and then focus on the SA sub-problem based on the characterization of the updated PA variables.

\vspace{-1em}
\subsection{Power Allocation Optimization (PA-Routine)}\label{sec:PA_routine}
For the PA sub-problem with a particular SA choice $\tilde{\qa}$, the optimization problem $\mathcal{P}_2$ reduces to a less complex problem with an affine objective function $P_C(\OID\!\!,\OEH\!\!,\tilde{\qa})$ subject to second order conic (SOC) constraints. We propose an optimization algorithm to obtain primal-dual feasible solutions using DR splitting based ADMM. In general, this mathematical operator is used to solve optimization problems of the form \cite{pontus}:
\begin{equation}
    \min_{\bm{x}} \big\{f(\bm{x})+g(\bm{x})\big\},\nonumber
\end{equation}
where $\bm{x}$ is the set of optimization variables, while $f$ and $g$ are proper closed and convex functions. This approach is applied on the problem $\mathcal{P}_2$ with $\bm{x} \rightarrow [\OID\!\!,\OEH]$, where $f$ corresponds to the affine objective function in \eqref{eq:objective_function_2} and $g$ relates to the problem constraints, including SOC conditions. From any initial $z^0 \in \mathbb{R}$ for iteration ($u$) of the PA procedure, the following DR splitting based primal-dual iterative ADMM updates are calculated for the specific case of $\alpha\!=\!1/2$ ~\cite[Eq. (25)-(28)]{pontus}.
\vspace{-0.2em}
\begin{subequations}\label{eq:ADMM_updates}
    \begin{align}
        \bm{x}^{\!(u+1)}\!\!&=\!\argmin_{\bm{x}}  \!\Big\{\! f(\bm{x}^{\!(u)})\!\!+\! 2 \tau \big \lVert A (\bm{x}^{\!(u)})\!\!+\!\! B(\bm{y}^{(u)}) \!\!-\!\! \qc \!+\! \bm{z}^{(u)}\big \rVert^2_2\!\Big\},\\
        \bm{x}_A^{(u+1)}\!\!&=\! 2 \alpha A (\bm{x}^{(u+1)}) -(1-2 \alpha)(B(\bm{y}^{(u)}) -\qc),\\
        \bm{y}^{(u+1)}\!\!&=\!\argmin_{\bm{y}} \! \Big\{\! g(\bm{y}^{\!(u)}) \!+\! \dfrac{\tau}{2} \big\lVert \bm{x}_A^{\!(u+1)} \!+\! B(\bm{y}^{\!(u)}) \!-\! \qc \!+\! \bm{z}^{(u)} \big\rVert^2_2 \Big\}, \\
        \bm{z}^{\!(u+1)}\!\!&= \!\bm{z}^{\!(u)} + \big(\bm{x}_A^{\!(u+1)} + B(\bm{y}^{\!(u+1)})-\qc \big), 
    \end{align}
\end{subequations}
where
\begin{align*}
     \!\!A(\bm{x})\!\!=\!\!\! \left[\!\!\!\! 
        \begin{array}{c}
        \,\,\lVert  \Psi^{\mathsf{\, nc}}_{\ID,l}(\OID\!\!,\OEH\!\!,\tilde{\qa}) \rVert_2 \!-\! \Xi(R^{\mathsf{EA}}_l)\lVert \Psi^{\mathsf{\, c}}_{\ID,l}(\OID\!\!,\tilde{\qa})\rVert_2\\
        \Lambda (I^{\mathsf{NL\!-\!EA}}_m)-\lVert I_m(\OID\!\!,\OEH\!\!,\tilde{\qa})\rVert_2\\
        \sus \ats \left(\suml  \osl + \summ  \osm \right) - P_t \\
        \suml \osl + \summ  \osm  - P_{s}
        \end{array} 
        \!\!\!\!\right],
\end{align*}

\vspace{0em}

\hspace{-1em}while $\qc\!=\!\left[\Xi(R^{\mathsf{EA}}_l), \Lambda (I^{\mathsf{NL\!-\!EA}}_m), P_t, P_s\right]^T$, and $B(\bm{y})=\Pi_{\mathcal{K}_{\text{SOC}}}(\qc-A(\bm{x}))$ defines a set of slack variables, which is introduced to relax the feasibility constraints using the projection operator. This operator $\Pi_{\mathcal{K}_{\text{SOC}}}(q)$, where $q = (q_0,q_1) \in \mathbb{R} \times \mathbb{R}^n$ is defined as the closest point on the second-order cone $\mathcal{K}_{\text{SOC}} = \left\{ (r, s) \in \mathbb{R} \times \mathbb{R}^n \mid \| s \|_2 \leq r \right\} \forall \, r \in \mathbb{R}, s \in \mathbb{R}^n$ to the point $q$. Mathematically, this can be represented as
\begin{equation}
    \Pi_{\mathcal{K}_{\text{SOC}}}(q) = \argmin_{(r, s) \in \mathcal{K}_{\text{SOC}}} \| (r, s) - (q_0, q_1) \|_2.
\end{equation}

At each iteration $u$, the algorithm checks for the convergence condition: $$\lvert P^{(u)}_C(\tOmgIDt\!,\tOmgEHt\!,\tilde{\qa}^{(t)})\!-\! P^{(u\!-\!1)}_C(\tOmgIDt\!,\tOmgEHt\!,\tilde{\qa}^{(t)}) \rvert \!\leq\! \epsilon,$$ for $\epsilon >0$. If this inequality condition is satisfied, the PA routine is terminated with the output PA estimates ($\tOmgID$\!,\! $\tOmgEH$) for the parametrized SA choice $\tilde{\qa}$.


\begin{figure*}[t]
    \centering
    \begin{minipage}[t]{0.32\textwidth}
        \centering
        \includegraphics[width=1.05\textwidth]{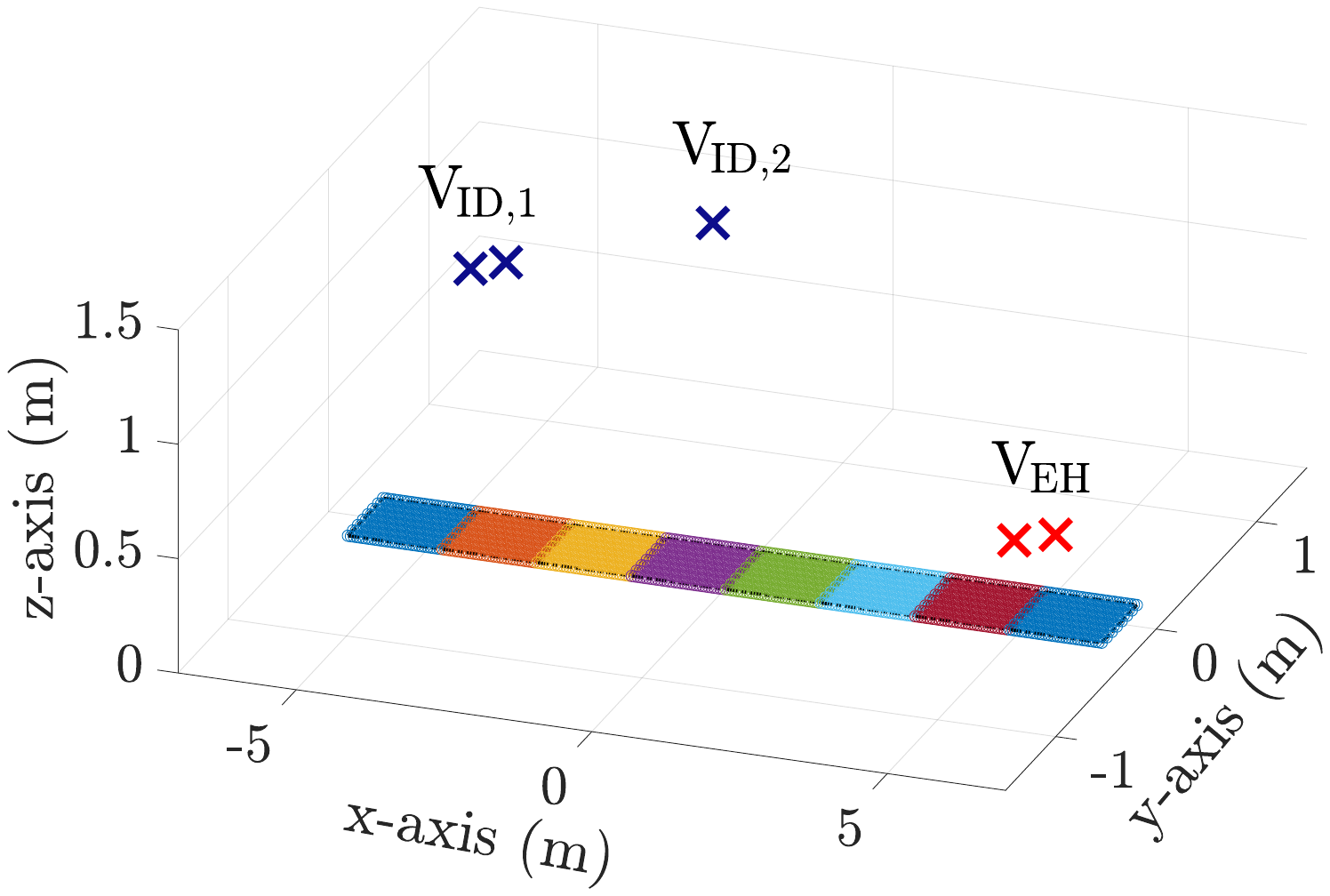}
        \vspace{0em}
        \caption{\small XL-MIMO system with $8$ subarrays of $32 \times 8$ elements with $3$ ID and $2$ EH users.\normalsize}
        \label{fig:ELAA_SWIPT}
    \end{minipage}
    \hfill
    \begin{minipage}[t]{0.66\textwidth}  
        \centering
        \begin{subfigure}[t]{0.475\textwidth}
            \centering
            \includegraphics[width=1.05\textwidth]{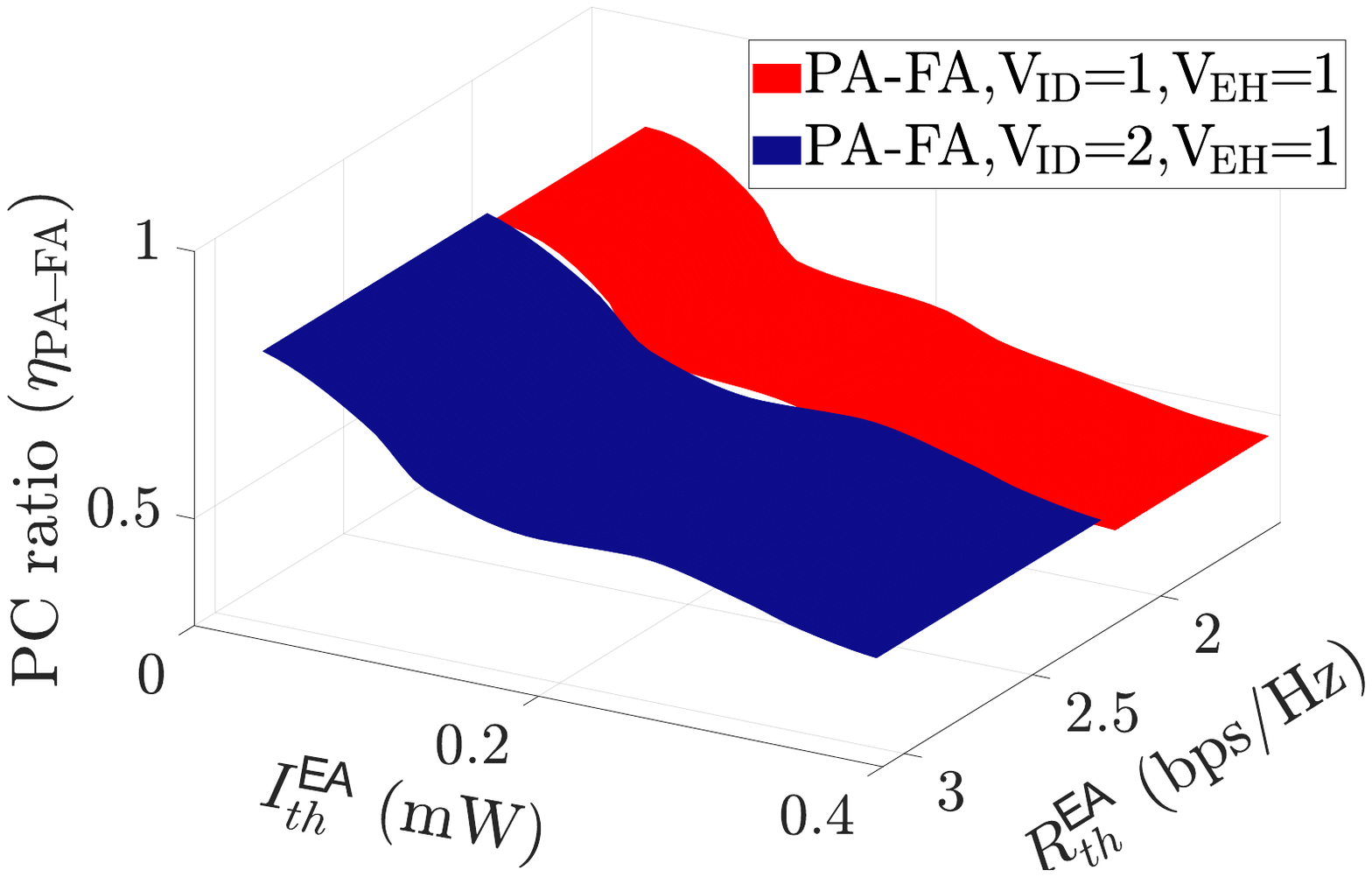}
            \caption{PA-FA case}
            \label{fig:P_C_thresholds_S4_V1}
        \end{subfigure}
        \hfill
        \begin{subfigure}[t]{0.475\textwidth}
            \centering
            \includegraphics[width=1.05\textwidth]{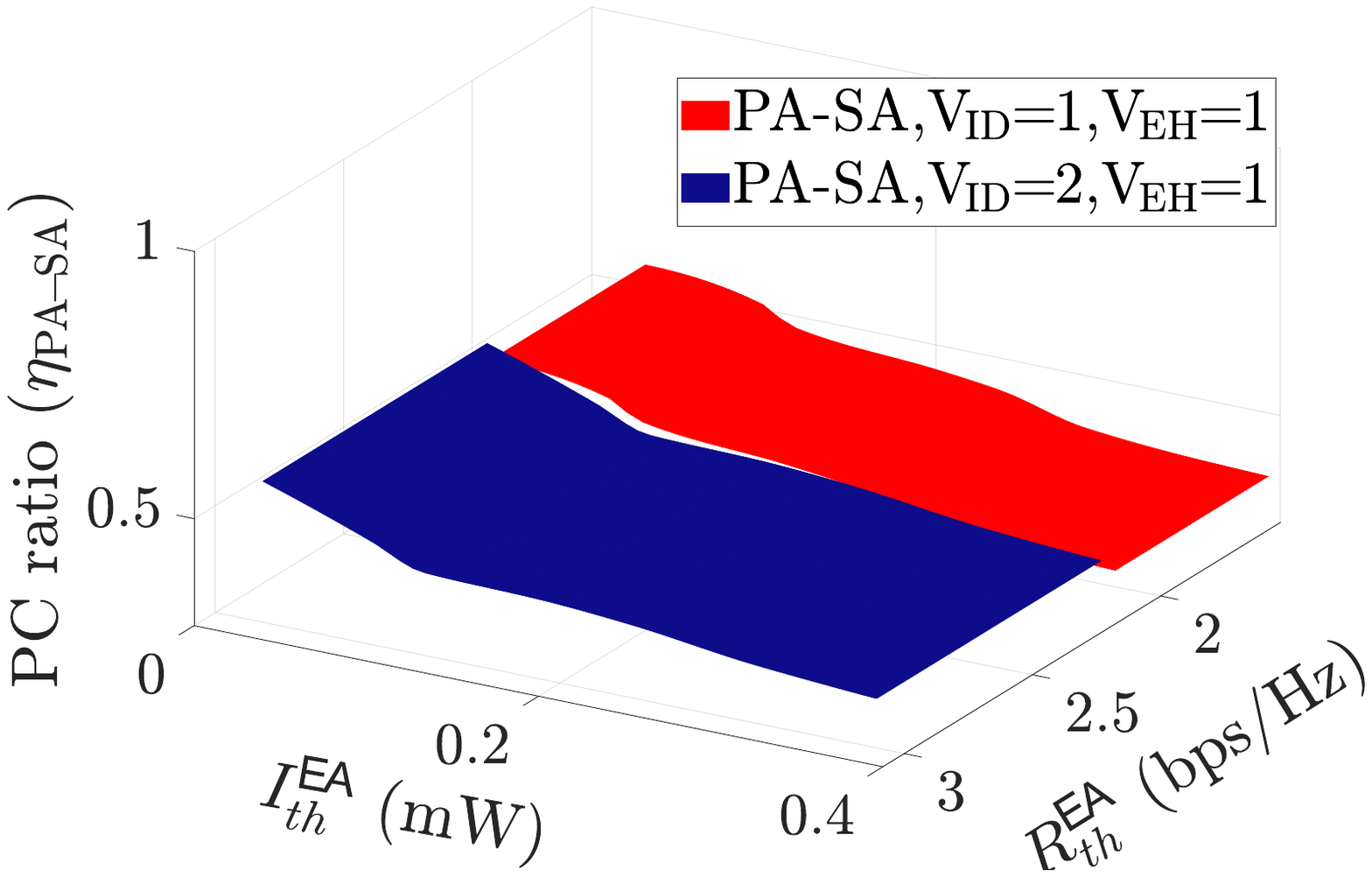}
            \caption{PA-SA case}
            \label{fig:P_C_thresholds_S4_V2}
        \end{subfigure}
        \vspace{-0.15cm}
        \caption{\small PC ratio over DL rate $R^{\mathsf{EA}}_{th}$ and EH power threshold $I^{\mathsf{EA}}_{th}$ for the $S=4$ subarray configuration. \normalsize}
        \label{fig:combined_thresholds}
    \end{minipage}
\vspace{-1.3em}
\end{figure*}
\vspace{-0.5em}
\subsection{Subarray Activation Optimization (SA Routine)}
Now, we discuss the SA routine, which evaluates the SA variables ($\qa$) for effective subarray activation using the estimated PA variables based on the PA routine delineated in Section \ref{sec:PA_routine}. Within a certain iteration $(t)$ of the SA routine, the PA estimates ($\tOmgID$\!,\! $\tOmgEH$) are calculated for a particular subarray activation choice ($\qa$), with initial realization of the FA setup $\qa =\boldsymbol{1}_S$. To this end, we propose a joint surrogate auxiliary function $h^t_s(\tOmgID\!\!,\tOmgEH)$, which incorporates the relative contribution of each active subarray in providing both DL ID and EH services to the network users, while satisfying the QoS requirements in PA optimization routine. This is mathematically defined as
\vspace{-0.3em}
\begin{align}~\label{eq:relative_contribution}
     \!\!\! h^t_s (\tOmgIDt\!,\tOmgEHt)\!\! = \!\! \dfrac{\varrho^{(t)} \suml \tilde{\Omega}^{\mathsf{ID\!,(t)}}_{s,l} \!+\! \summ \tilde{\Omega}^{\mathsf{EH,(t)}}_{s,m}\!}{\sus\big(\varrho^{(t)} \suml \tilde{\Omega}^{\mathsf{ID\!,(t)}}_{s,l} \!\!+\!\! \summ \tilde{\Omega}^{\mathsf{EH,(t)}}_{s,m}\big)}\!,\!
     \vspace{-0.7em}
\end{align}
where $\varrho^{(t)}=\sus \summ \tilde{\Omega}^{\mathsf{EH,(t)}}_{s,m}/\sus \suml \tilde{\Omega}^{\mathsf{ID,(t)}}_{s,l}$ is the balancing factor which adjusts the weight of the power required for supporting the ID QoS conditions in correlation with the power allocated for EH operations. The  introduction of this parameter in the surrogate function allows us to access the overall performance of each XL-MIMO subarray in SWIPT operations over two distinct functional domains (ID and EH) which have entirely different power threshold requirements. This corresponds to the fact that the function $h^t_s(\tOmgID\!,\tOmgEH)$ yields higher value for the subarray which has a stronger contribution towards the overall system performance in comparison to the other subarrays in the XL-MIMO array, and vice versa.

\begin{algorithm}[t]
\caption{Joint Optimization of Subarray Activation ($\qa$) and SWIPT Power Allocation ($\OID \!,\OEH$)}\label{alg:opt_process}
\begin{algorithmic}[1]

\State \textbf{Initialize:} Set iteration indices $t = 0$ and $u = 0$, convergence thresholds $\delta\!>\!0$ and $\epsilon\!>\!0$.
\State Initial estimates: $\tilde{\Omega}^{(0)}_{s,l}=\tilde{\Omega}^{(0)}_{s,m}=P_s/K$, $\tilde{a}^{0}_s=1$
\State \textbf{SA Routine:}
\Repeat

    \State Compute the surrogate auxiliary function
    using~\eqref{eq:relative_contribution}.
    
    \State Update the subarray activation variable $a^t_s$ using \eqref{eq:binary_decision}.
    \State Scale the activation variable $\ats$ using \eqref{eq:parameterized}.
    \State Formulate the parameterized optimization problem:
    \begin{equation}\label{eq:para_opt}
        \min_{\{\OID,\OEH\}} \,\,P_C(\tOmgIDt\!,\tOmgEHt,\tilde{\qa}^{(t)}) \nonumber
    \end{equation}
    \State\textbf{PA Routine:}
    \Repeat
        \State Form the conic representation using~\eqref{eq:optimization_2}.
        \State Solve the 
        parametrized optimization problem in Step 8 using DR splitting based ADMM in \eqref{eq:ADMM_updates} with $u$ sub-iteration.
        \State Increment inner iteration index: $u = u + 1$.
        \Until{$\lvert P^{(u)}_C(\tOmgIDt\!,\tOmgEHt\!,\tilde{\qa}^{(t)})\!-\! P^{(u\!-\!1)}_C(\tOmgIDt\!,\tOmgEHt\!,\tilde{\qa}^{(t)}) \rvert \!\leq\! \epsilon$}
    
    \State Increment outer iteration index: $t = t + 1$.
    \State Update PA estimates $\tilde{\boldsymbol{\Omega}}^{\mathsf{ID,t}}$\!,\! $\tOmgEHt$
    
\Until{$ \lvert P_C(\tOmgIDt\!\!,\tOmgEHt\!,\tilde{\qa}^{(t)})\!-\! P_C(\tilde{\boldsymbol{\Omega}}^{\mathsf{ID, (t\!-\!1)}}\!\!,\tilde{\boldsymbol{\Omega}}^{\mathsf{EH,(t\!-\!1)}}\!\!,\tilde{\qa}^{(t\!-\!1)}) \rvert \!\leq\! \delta$}
\State Return optimized parameters as $\OmgIDst=\tilde{\boldsymbol{\Omega}}^{\mathsf{ID, (t)}}$, $\OmgEHst=\tilde{\boldsymbol{\Omega}}^{\mathsf{EH,(t)}}$ and $\qa^{\star}=\tilde{\qa}^{(t)}$.
\end{algorithmic}
\end{algorithm}
\setlength{\textfloatsep}{0.3cm}

Next, we associate the relative contribution based surrogate function $h^t_s(\tOmgID\!,\tOmgEH)$ to the binary decision SA variables $\qa$. In the iteration ($t$), we activate the subarrays which have higher $h^t_s(\tOmgID\!,\tOmgEH)$ than the mean threshold $\ME\{h^t_s(\tOmgID\!,\tOmgEH)\}$ and switch off the other subarrays. This iterative process provides the set of SA estimates $\tilde{\qa}$ whose individual elements are mathematically represented as:
\begin{align}
\label{eq:binary_decision}
     a^t_s \!&=\! 
        \begin{cases}
            1 \quad \,\, h^t_s(\tOmgIDt\!,\tOmgEHt)\! \geq \! \ME\{h^t_s(\tOmgIDt\!,\tOmgEHt)\},\\
            0 \quad  \,\, h^t_s(\tOmgIDt\!,\tOmgEHt)\! < \!\ME\{h^t_s(\tOmgIDt\!,\tOmgEHt)\}.
        \end{cases}
\end{align}

The mean threshold $\ME\{h^t_s(\tOmgID\!,\tOmgEH)\} (1/S)$ assures that only those subarrays, which have a significant role in the SWIPT performance, are activated. At this stage, these updated SA variables are scaled once again with the same joint surrogate function defined in \eqref{eq:relative_contribution} for the next iteration ($u+1$) of the PA routine. Now, the continuous SA estimates are
\begin{equation}\label{eq:parameterized}
     \tilde{a}^t_s = h^t_s(\tOmgIDt\!,\tOmgEHt) a^t_s.
\end{equation}

This scaling process maintains the relative magnitude of performance contribution within the subset of activated subarrays. The parametrized SA estimates $\tilde{\qa}^{(t)}$ are substituted in \eqref{eq:ADMM_updates}, thus, triggering the next iterative cycle of PA primal-dual updates. 

\vspace{-0.5em}
\subsection{Overall Algorithm and Complexity Analysis} 
The PA and SA procedures are implemented alternatively, as detailed in \textbf{Algorithm \ref{alg:opt_process}}. For a given set of PA coefficients, the array switching vector $\qa$ is optimized. Then, given $\qa$, the PA coefficients are optimized until the optimization convergence is reached within a negligible tolerance, i.e., $ \lvert P_C(\tilde{\boldsymbol{\Omega}}^{\mathsf{ID ,t}}\!,\tOmgEHt\!,\tilde{\qa}^{(t)})\!-\! P_C(\tilde{\boldsymbol{\Omega}}^{\mathsf{ID, (t\!-\!1)}}\!,\tilde{\boldsymbol{\Omega}}^{\mathsf{EH,(t\!-\!1)}}\!,\tilde{\qa}^{(t\!-\!1)}) \rvert \!\leq\! \delta$ for $\delta >0$.

The execution of the PA routine relies on the proximal operators within each DR based ADMM iteration, incurring a computational complexity of $\mathcal{O}(SK)$ for each subarray-user pair. Furthermore, the convergence rate for this sub-routine scales linearly with the problem size $\mathcal{O}(SK)$. In the SA procedure, the evaluation of the surrogate function requires computations across $SK$ power coefficients. Additionally, the binary decision-making process in \eqref{eq:binary_decision} over $\ME\{h^t_s(\tOmgIDt\!,\tOmgEHt)\}$ exhibits a logarithmic scaling of $\mathcal{O}(\log(SK))$. Consequently, the total computational complexity of the proposed PA-SA algorithm is $\mathcal{O}(S^2K^2\log(SK))$.

\vspace{0em}
\section{Numerical Results}
\begin{figure*}[t]
\vspace{-0.5cm}
    \centering
    \begin{minipage}[t]{0.32\textwidth}
        \centering
        \includegraphics[trim=0 380 0 0,clip,width=1.12\textwidth]{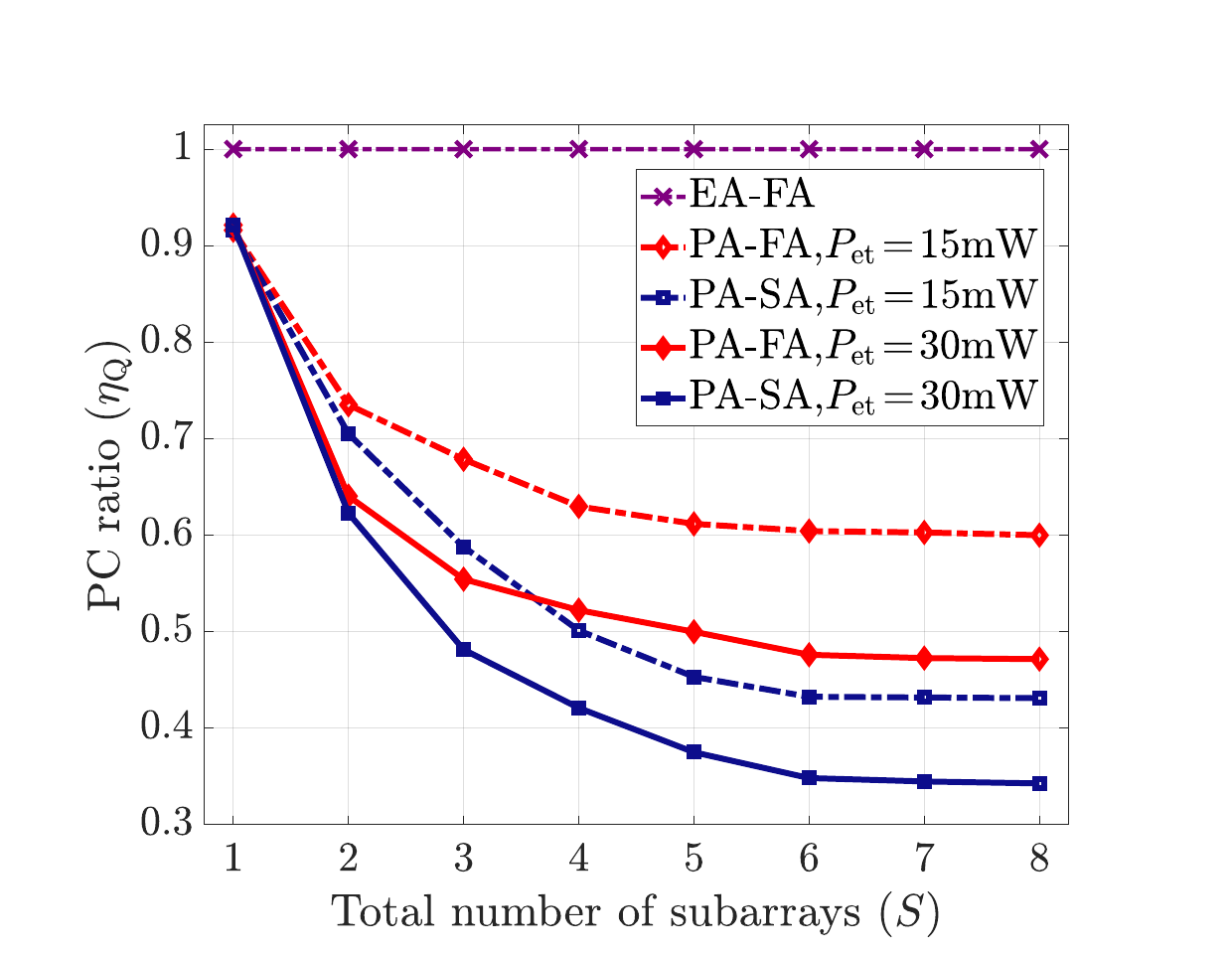}
        \caption{\small XL-MIMO PC ratio vs the number of subarrays ($\mathrm{{V}_{\!ID}}\!=\!1$, $\mathrm{{V}_{\!EH}}\!=\!1$).\normalsize}
        \vspace{-2cm}
    \label{fig:power_consumption_ratio_V1}
    \end{minipage}
    \hfill
    \begin{minipage}[t]{0.32\textwidth}
        \centering
        \includegraphics[trim=0 1.4cm 0cm 0cm,clip,width=1.28\textwidth]{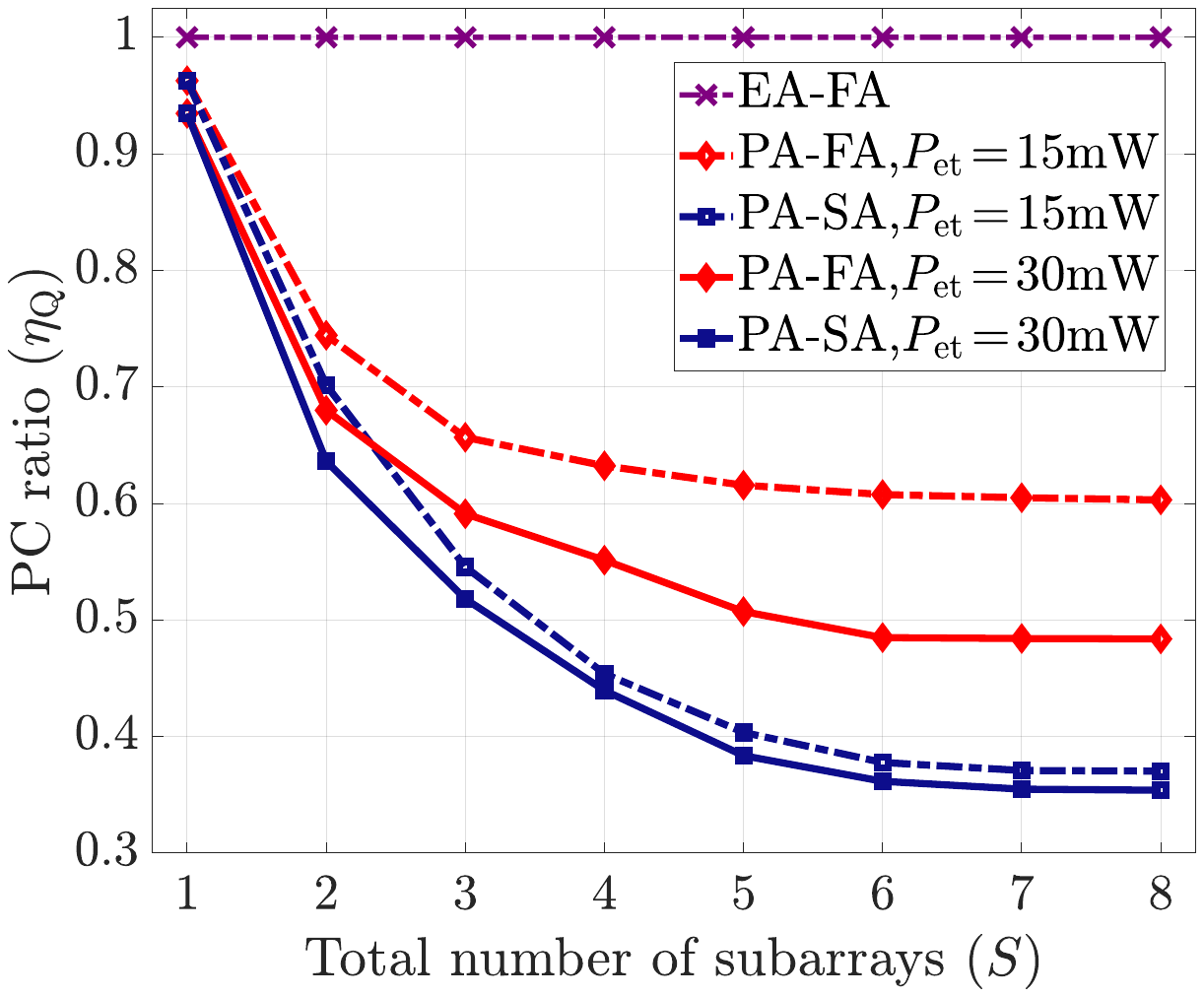}
        \caption{\small XL-MIMO PC ratio vs the number of subarrays ($\mathrm{{V}_{\!ID}}\!=\!2$, $\mathrm{{V}_{\!EH}}\!=\!1$).\normalsize}
        \label{fig:power_consumption_ratio_V2}
    \end{minipage}
    \hfill
    \begin{minipage}[t]{0.32\textwidth}
        \centering
        \includegraphics[trim=0 1.4cm 0cm 0cm,clip,width=1.28\textwidth]{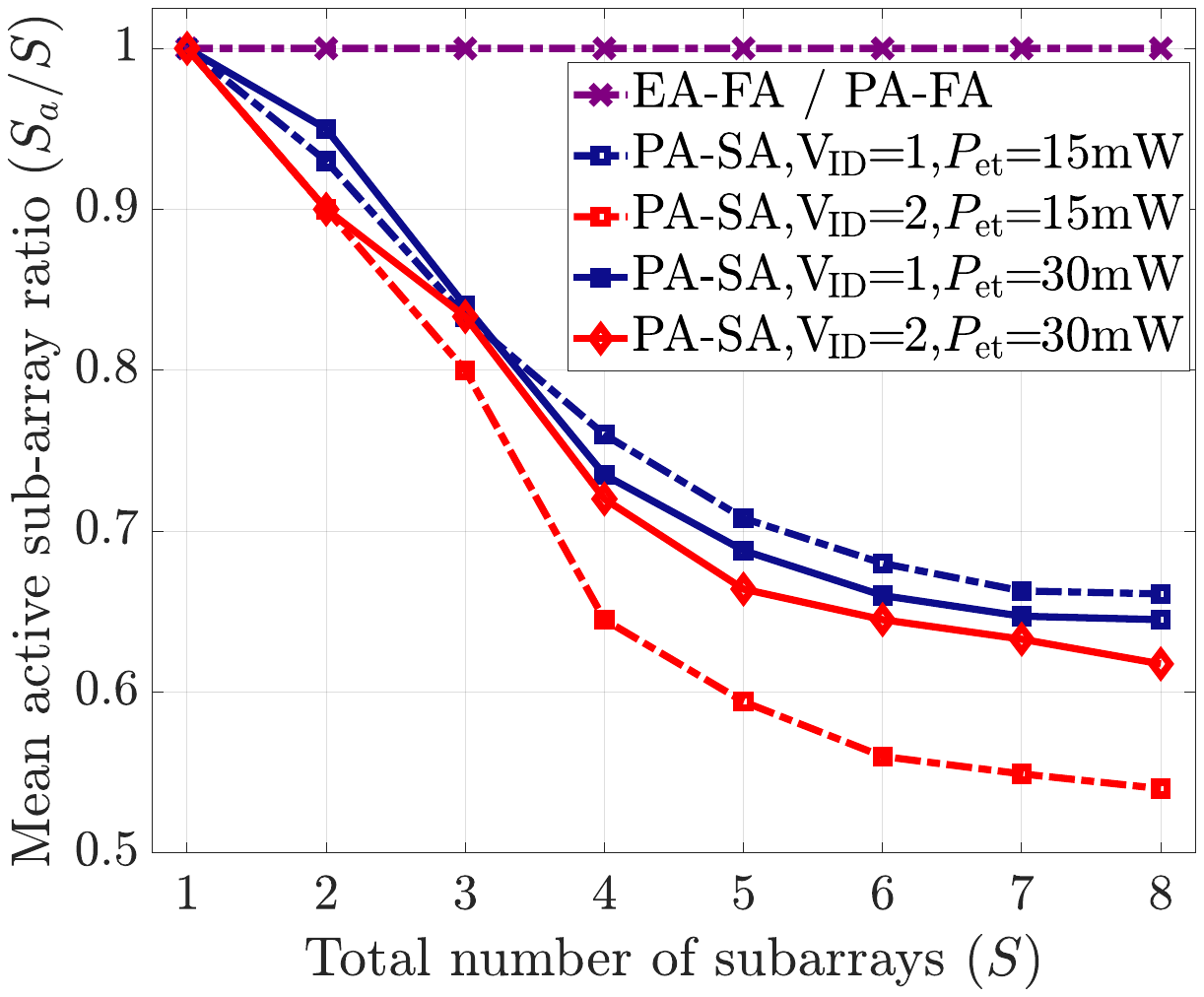}
        \caption{\small Mean active subarray ratio vs the number of subarrays.\normalsize}
        \label{fig:Active_subarray_ratio}
    \end{minipage}
\vspace{-1em}
\end{figure*}

In this section, we examine the performance of the considered XL-MIMO system in Fig. \ref{fig:ELAA_SWIPT} with the proposed joint optimization scheme, where the transmission array comprises $S$ subarrays, each equipped with $\ns=256$ antennas. The antennas in these UPSAs are arranged as $N_{s,x}=32$ elements along the $x$-axis and $N_{s,y}=8$ elements along the $y$-axis. Each antenna element has the largest physical dimension as $D=\lambda/4$ with the system carrier wavelength $\lambda=0.1$ m \cite{Ramezani2024}. The XL-MIMO system is intended to serve a total $K=5$ users with $L=3$ ID and $M=2$ EH users. The EH users are located in a single VR, while the ID users are situated in two distinct VRs, $\mathrm{{V}_{\!ID}}=\{1,2\}$. The ID users are limited within the range of $d_{FA}/10$ from the XL-MIMO array and EH users within $d_{FA}/100$ \cite{Zhang:JSAC:2024}. Unless stated otherwise, the PC settings are adopted from \cite{Jun_Zhang} and are given as $\varsigma=0.35$, $P_{\mathrm{syn}}=50$ mW, $P_{\mathrm{ct}}=48.2$ mW, and $P_{\mathrm{et}}=30$ mW. The noise power for each ID user is $\sigma_l=-80$ dBm. The parameters for the NL-EH model are chosen as $\zeta_{max}=24$ mW and circuit parameters are $a=1500$, $b=0.0022$ \cite{Nezhadmohammad}. 
The algorithm convergence tolerance parameters are set to $\epsilon=\delta=10^{-7}$.

To evaluate the effectiveness of the proposed optimization framework, we compare three methods: \textit{i)}  Equal power allocation with ${\boldsymbol{\Omega}}^{\mathsf{ID}}\!\!=\!{\boldsymbol{\Omega}}^{\mathsf{EH}}\!\!=\!\! P_s/K$ and full array activation, i.e.,  $\qa=\boldsymbol{1}_S$ (\textbf{EA-FA}); \textit{ii)}  Optimized power allocation $\OmgIDst$, $\OmgEHst$ with full array activation, i.e.,  $\qa=\boldsymbol{1}_S$ (\textbf{PA-FA}); and \textit{iii)}   Proposed joint optimization scheme with PA-procedure based optimized PA $\OmgIDst$, $\OmgEHst$ and SA $\qa^{\star}$ (\textbf{PA-SA}).

First, we analyze the performance of the above-mentioned methods in terms of the  overall system PC, $P_C$, and the transmitted power $P_{T\!X}= \sus (\suml  \osl \! +\! \summ\!  \osm)$ by the XL-MIMO array. The results for these parameters are presented in Table~\ref{tab:power_consumption}. We observe that the overall $P_C$ levels for the \textbf{PA-SA} case are up to $65\%$ lower than those for the \textbf{EA-FA} case and $30\%$ lower than for the \textbf{PA-FA} case. Moreover, the $P_{T\!X}$ values for both the \textbf{PA-SA} and \textbf{PA-FA} cases are up to $80\%$ lower than those for the \textbf{EA-FA} case. Although the $P_{T\!X}$ levels for \textbf{PA-SA} are similar to those for \textbf{PA-FA}, there is a considerable difference in the $P_C$ levels between these two cases. This observation highlights the impact of SA, where the optimized $P_{T\!X}$ is primarily allocated to a smaller set of subarrays to meet the user QoS requirements, thereby significantly reducing the circuitry power dissipation by deactivating the remaining subarrays.

\begin{table}[t]
\centering
    \caption{Comparison of PC, $P_C$ and transmitted power, $P_{TX}$ for scenario $\mathrm{{V}_{\!ID}}=2$, $\mathrm{{V}_{\!EH}}=1$}
    \begin{tabular}{c | *{3}{S[table-format=3.2]} | *{3}{S[table-format=3.2]}}
    \toprule
     & \multicolumn{3}{c|}{\textbf{Power Consumption $\boldsymbol{P_C}$(W)}} & \multicolumn{3}{c}{\textbf{Transmitted Power $\boldsymbol{P_{T\!X}}$(W)}} \\
    \textbf{S} & \textbf{EA-FA} & \textbf{PA-FA} & \textbf{PA-SA} & \textbf{EA-FA} & \textbf{PA-FA} & \textbf{PA-SA} \\
    \midrule
    \textbf{1} & 34.38205714 &	32.13539198	& 32.13539198
    & 7.68 &	6.893667194 &	6.893667194\\
    \textbf{2} & 68.76411429 &	46.75303697	& 43.744867
    & 15.36 &	7.656122939	& 7.402926
     \\
    \textbf{3} & 103.1461714 &	60.96860224 & 53.4356
     & 23.04	& 8.277850785	& 8.378734886\\
    \textbf{4} & 137.5282286 &	75.81445431 & 60.44190959
     & 30.72	& 9.12017901 & 8.615954758\\
    \textbf{5} & 171.9102857 &	87.18959529	& 65.93084237
     & 38.4	& 9.747758351 &
     9.621444431 \\
    \textbf{6} & 206.2923429 &	100.0129 &	74.5774
    & 46.08	& 10.32625129 & 10.56661675 \\
    \textbf{7} & 240.6744 &	116.4970276 & 85.3844337
     & 53.76	& 10.99791966 &	10.8928118 \\
    \textbf{8} & 275.0564571 &	133.045721 &	95.6776
     & 61.44	& 11.73624236 & 11.49842456 \\
    \bottomrule
    \end{tabular}
    \label{tab:power_consumption}
\end{table}

To gain deeper insights into the comparative performance improvements, the PC ratio is defined with respect to the benchmark \textbf{EA-FA} case as $\eta_Q \triangleq P^{Q}_C / P^{\text{EA-FA}}_C$, where $Q \in \{\text{\textbf{PA-SA, PA-FA, EA-FA}}\}$. This parameter is analyzed over the QoS thresholds for the DL rate $R^{\mathsf{EA}}_{th}$ and EH power $I^{\mathsf{EA}}_{th}$ for a specific configuration of $S=4$ subarrays with two ID VR cases: 1) $\mathrm{{V}_{\!ID}}=1$ in Fig. 2a and 2) $\mathrm{{V}_{\!ID}}=2$ in Fig. 2b. It can be noticed that $\eta_Q$ decreases for both \textbf{PA-SA} and \textbf{PA-FA} methods with increasing $I^{\mathsf{EA}}_{th}$ (from $0.04$mW to $0.4$mW), which means that the proposed optimization design provides much lower $P^{Q}_C$ in comparison to \textbf{EA-FA}. Specifically, we observe that for the $\mathrm{{V}_{\!ID}}=1$ case, $\eta_{\text{PA-FA}}$ decreases from $0.81$ to $0.46$, and $\eta_{\text{PA-SA}}$ decreases from $0.56$ to $0.39$. Similarly, for the $\mathrm{{V}_{\!ID}}=2$ case, $\eta_{\text{PA-FA}}$ reduces from $0.83$ to $0.50$, while $\eta_{\text{PA-SA}}$ decreases from $0.59$ to $0.42$. This relates to the fact that the reduction in $P^{\text{PA-SA}}_C$ for the proposed method is up to $61\%$ against the $P^{\text{EA-FA}}_C$ and $30\%$ in comparison to $P^{\text{PA-FA}}_C$. We also observe that $P^Q_C$ remains unaffected within the range of the DL rate threshold $R^{\mathsf{EA}}_{th}$, since an increase in the transmitted power to meet higher $R^{\mathsf{EA}}_{th}$ also generates more signal interference, particularly in the \textbf{EA-FA} case, which serves as the baseline method. However, the VR scenario impacts the positioning of users, thereby influencing inter-user interference. As a result, we observe higher $R^{\mathsf{EA}}_{th}$ values for the $\mathrm{{V}_{\!ID}}=2$ case, which is associated with reduced interference from inter-ID-VR users.

We now examine the decreasing trends of $\eta_Q$ over the increasing number of subarrays ($S$) for $\mathrm{{V}_{\!ID}}=1,2$, as shown in Fig. \ref{fig:power_consumption_ratio_V1} and \ref{fig:power_consumption_ratio_V2}, respectively. For both scenarios with $P_{\mathrm{et}}=15$, and $30$mW, the $\eta_{\text{PA-SA}}$ values are reduced up to $65\%$ against $\eta_{\text{EA-FA}}$ and $32\%$ in comparison to $\eta_{\text{PA-FA}}$, which can be attributed to more concentrated VRs with an increasing physical size of modular XL-array. To fully comprehend this effect while considering the non-stationary channels, the mean ratio of active subarrays, denoted by $S_a/S$, is an important parameter to evaluate the performance of SA procedure. The results presented in Fig. \ref{fig:Active_subarray_ratio} substantiate the fact that a smaller number of subarrays  (i.e., $\sim 60\%$)  can predominantly support the SWIPT operation with expanding physical dimensions of the XL-MIMO array and more clustered users' spatial positions.

\vspace{0em}
\section{Conclusion}
We investigated a practical SWIPT framework using a modular XL-MIMO system in the near-field communication domain, while accounting for spatial non-stationarities. In this context, we minimized the overall system PC by jointly optimizing the PA for both ID and EH users along with SA, subject to the QoS requirements of DL ID rate and EH power constraints. The proposed two-layer optimization algorithm decouples the mixed-integer problem into PA and SA based sub-problems, and resolves them utilizing instrumental techniques, such as DR splitting based ADMM method and surrogate auxiliary function. The numerical results validated the significant power conservation capability of our proposed method, achieved by employing fewer active subarrays. Future research could examine VR effects in mixed near-/far-field SWIPT with hybrid beamforming for better energy efficiency.


\vspace{-0.5em}
\bibliographystyle{IEEEtran}
\bibliography{main}

\end{document}